 \documentclass[aps,prb,floatfix,onecolumn]{revtex4}
\usepackage[english]{babel}

\usepackage{graphicx}
\usepackage{dcolumn}
\usepackage{bm}
\usepackage{epstopdf}

\newcommand{\rot}{{\rm rot}}

\newcommand{\myskip}[1]{}

\renewcommand{\d}{{\rm d}}

\newcommand{\BEQ}{\begin{eqnarray}}
\newcommand{\EEQ}{\end{eqnarray}}
\newcommand{\BEA}{\begin{eqnarray}}
\newcommand{\EEA}{\end{eqnarray}}
\newcommand{\nn}{\nonumber}
\newcommand{\Sigmab}{\overline\Sigma}

\newcommand{\m}{{\rm m}}
\newcommand{\cm}{{\rm cm}}
\newcommand{\gr}{{\rm gr}}

\newcommand{\s}{{\rm s}}

\newcommand{\kpc}{{\rm kpc}}
\newcommand{\Mpc}{{\rm Mpc}}
\newcommand{\eV}{{\rm eV}}

\newcommand{\cg}{{\it cg}}
\newcommand{\co}{{\it co}}

\newcommand{\Gal}{{\it G}}

\newcommand{\vp}{{\bf p}}
\newcommand{\vv}{{\bf v}}

\newcommand{\LCDM }{$\Lambda$CDM}

\newcommand{\half}{\frac{1}{2}}

\begin{document}

\title{How Zwicky already ruled out modified gravity theories without dark matter}

\author{Theodorus Maria Nieuwenhuizen$^{1,2}$}

\affiliation{$^1$Institute for Theoretical Physics, University of Amsterdam, Science Park 904, 
 1090 GL  Amsterdam, The Netherlands \\ $^2$International Institute of Physics, UFRG, 
Lagoa Nova, Natal - RN, 59064-741, Brazil}



\begin{abstract}
Various theories, such as MOND, MOG, Emergent Gravity and $f(R)$ theories
avoid dark matter by assuming a change in General Relativity and/or in Newton's law.
Galactic rotation curves are typically described well. 
Here the application to galaxy clusters is considered, focussed on the good lensing and X-ray data for A1689.
 As a start, the no-dark-matter case is confirmed to work badly: 
 the need for dark matter starts near the cluster centre, where Newton's law is still supposed to be valid.
 This leads to the conundrum discovered by Zwicky, which is likely only solvable in his way, namely by assuming additional (dark) matter. 
 Neutrinos with eV masses serve well without altering the successes in (dwarf) galaxies.
\end{abstract}


\maketitle


\section{Introduction}

\hfill{\it  Niets is gewichtiger dan donkere materie}\footnote{Nothing matters more than dark matter}

\vspace{3mm}

{\it Dark matter} enters the light in 1922 when Jacobus Kapteyn employs this term 
in his {\it First Attempt at a Theory of the Arrangement and Motion of the Sidereal System} \cite{kapteyn1922first}. 
It took a decade for his student Jan Oort to work out the stellar motion perpendicular to the Galactic plane, and to conclude that 
{\it invisible mass} should exist to keep the stars bound to the plane \cite{oort1932force}. 
Next year, in 1933, Fritz Zwicky applied the idea to the Coma galaxy cluster and concluded that it is held together by dark matter. Interestingly, 
he starts with the expression {\it dunkele (kalte) Materie} [(cold) dark matter, (C)DM] \cite{zwicky1933rotverschiebung},
which is now popular but just referred to cool and cold stars, macroscopic and microscopic bodies and gases \cite{zwicky1937masses}.
In principle, the galactic and cluster DM can have a different cause, so we may speak of
  {\it Oort DM}, baryonic or not, and nonbaryonic {\it Zwicky DM}.
  
Presently, it is assumed that both types of DM coincide.
The standard model of cosmology \LCDM \  is very successful in explaining the WMAP and Planck data \cite{hinshaw2013nine,planck2015planck}, 
so \LCDM \ is rightfully a good  {\it effective} theory. 
But is it a {\it fundamental} theory? Many puzzling observations make this nonevident. The CDM particle, 
the WIMP, if it exists, keeps on hiding itself 6 more years since its  ``moment of truth'' \cite{bertone2010moment}\footnote{Acronyms
of direct and indirect dark matter searches:
ADMX, AMS, AMANDA, ANAIS, ANTARES, ArDM, ATIC, BPRS, CANGAROO, CAST, 
CDEX, CDMS, Chandra, CLEAN, CoGeNT, COUPP, COSME, CRESST, CUORE, CYGNUS, 
DAMA/NaI, DAMA/LIBRA, DAMIC, DarkSide, DEAP, DM-Ice, DMTPC,  DRIFT, EDELWEISS, EGRET, 
ELEGANTS,  FERMI-LAT, GENIUS, GERDA, GEDEON, GLAST, HEAT, HESS, HDMS, ICECUBE, 
IGEX, INTEGRAL, KIMS, LEP, LHC, LUX, MAGIC, MALBEK, MIMAC, miniCLEAN, 
NaIAD, NEWAGE, ORPHEUS, PAMELA, Panda-X, PICASSO, PICO, ROSEBUD, SIMPLE, SUPER-K, 
Suzaku, TEXONO,  UKDMC, VERITAS, WArP, Whipple, XENON10/100, XMASS, XMM-Newton, ZEPLIN.  
Upcoming: ADMX-Gen2, ArDM-1t, CTA, DARWIN, EURECA, FUNK, GEODM, HAWC, LSST, LZ, MAJORANA,  SuperCDMS, DEAP-3600, XENON1T.}.
One observes 19 quasars with spins aligned with their hosts large-scale structures on a scale of almost 1 Gpc \cite{hutsemekers2014alignment}
and dozens of radiojets from AGNs aligned on a scale of 30 Mpc \cite{taylor2016alignments}.
 A ring (actually, a spiral) of 9 gamma ray bursts extends over nearly 2 Gpc \cite{balazs2015giant};
 the ``cosmic train wreck" galaxy cluster A520 has a central $3-4\,10^{13}M_\odot$ 
mass clump with mass-to-light ratio $800M_\odot/L_{R\odot}$ \cite{jee2012study,clowe2012dark,jee2014hubble,wang2016merging};
in the cluster A3827 the offset between baryonic and dark mass  \cite{massey2015behaviour}
is an order of magnitude `too large" \cite{schaller2015offsets}.  Puzzles in galaxies include: 
the brightness fluctuations in the Twin Quasar allow a DM interpretation in terms of a large halo of rogue planets 
in the lensing galaxy \cite{schild1996microlensing};
 the observed satellites of the Galaxy lie in a plane, not in random \LCDM \ directions \cite{kroupa2012failures};
the predicted transition for the most massive galaxies to transform from their initial halo assembly at redshifts $z=8-4$ 
to the later baryonic evolution known from star-forming galaxies and quasars is not observed \cite{steinhardt2015impossibly};
the galaxy power spectrum deduced from SDSS-III observations fits well to the stretched exponential $\exp[-(k/k_b)^{1/2}]$ 
from turbulence  \cite{bershadskii2015deterministic}.
Various further arguments can be found in our investigations \cite{nieuwenhuizen2009gravitational,nieuwenhuizen2010micro}.

Consequently, various studies consider other explanations for the DM effect. One option is that, beyond a certain scale, 
the force gets enhanced with respect to the Newton law.
The best known case is MOdified Newtonian Dynamics (MOND), where $g_m$,  the Newtonian gravitational acceleration due to the baryonic matter, 
gets replaced by $a=\sqrt{a_0g_m}$ in the domain $g_m<a_0$,
where the crossover acceleration is $a_0=cH_0/2\pi$ \cite{milgrom1983modification}.
In essence, this replaces the Newton fall of $g_m=GM/r^2$ by an effective fall off $a=\sqrt{a_0GM}/r$, which implies a flattening 
of the rotation curve, $v\to$ const for large $r$, as observed in galaxies. The approach acts as a faithful prediction for the
rotation velocity based on the knowledge of the baryonic matter, that is, stars, brown dwarfs, Jupiters, clouds, etc.
It achieves a scaling for a collection of rotation curves
and explains the baryonic Tully-Fisher relation $M\sim v^4_\rot$ \cite{famaey2012modified}.

Related deviations from Newton's law are MOG \cite{moffat2005gravitational} and $f(R)$ gravity
\cite{sotiriou2010f(R),defelice2010f(R)}, while Emergent Gravity \cite{verlinde2011origin} also leads to a variant of MOND.
These non-Newtonian gravities have mainly been used to explain Oort DM, while MOND is known to need neutrinos in 
clusters \cite{sanders2003clusters,sanders2007neutrinos}.

In recent years we have made a series of studies for the case of neutrinos as Zwicky DM, specialising to the cluster Abell 1689, for which
good lensing and gas data are available \cite{nieuwenhuizen2009non,nieuwenhuizen2011prediction,nieuwenhuizen2013observations}. 
This leads to a perfect fit without missing baryons for a neutrino mass of 1.9 eV, 
a case consistent with the 2 eV upperbound from tritium decay 
and, if cosmic structure formation is nonlinear, not excluded beforehand by neutrino free streaming
\cite{nieuwenhuizen2013observations,nieuwenhuizen2016dirac}.

We now use the A1689 data to investigate the alternative gravity theories in their application to Zwicky DM.
In section 2 we describe the data sets to be employed and in section 3 the quantities related to them.
In section 4 we analyse the Newton acceleration of the normal (baryonic) matter in comparison with estimates from the lensing data.
In the next sections we study in detail the application to MOND, MOG, Emergent Gravity and $f(R)$ theories, respectively. 
We close with a discussion.


\section{Data description}

Lensing properties of A1689 are expressed by strong lensing (SL), that is, background galaxies for which the ideal Einstein ring 
is partly achieved as a set of arclets, up to 5 of them appear to be connected to a common source galaxy.
Further information comes from weak lensing (WL): randomly oriented background galaxies get a systematic deformation (shear) 
which is distilled by averaging over the galaxies in a small field. Finally, X-ray observations yield the electron density and
thus the gas mass density. 

All our results are scaled to the flat \LCDM \  cosmology with  $\Omega_M= 0.3$,    $\Omega_\Lambda= 0.7$, 
and a  Hubble constant $H_0 = 70h_{70}$ km s$^{-1}$ Mpc$^{-1}$ with $h_{70}=1$.
At the redshift $z= 0.183$ of the A1689 cluster, $1''$ corresponds to 3.035 kpc. 

\subsection{Strong and weak lensing}

We collect data from the literature.
The SL analysis by Limousin et al. (2007) yields data for the line-of-sight mass density $\Sigma(r)$  
at radii  between 3 and 271 kpc from the cluster centre \cite{limousin2007combining}.  
The resulting 12 data points $(r_i,  \Sigma_i)$ and their correlation matrix were kindly supplied by M. Limousin;
see \cite{nieuwenhuizen2013observations} for the regulation of the small eigenvalues.
Essentially the same regime is covered by the SL analysis of Coe et al. (2010) \cite{coe2010high}, which has 20 data points;
these two data sets overlap in the window between 40 and 150 kpc with both having small errors, giving faith in the approaches.
However, discrepancies occur at small $r$: Coe et al. have 3 data points below 40 kpc while 
Limousin et al. have there 6 data points with smaller errors. Beyond 150 kpc the Coe data become increasingly noisy.
These differences stem from the different analysis and/or the neglect of correlations.
Umetsu et al. (2015) present 14 data points for $\Sigma$  between 125 kpc and 3 Mpc, derived mainly from WL  \cite{umetsu2015three},
and kindly supplied by K. Umetsu, with their correlation matrix.
These data have mostly substantially bigger errors than both previous sets, but are complementary and coincide within the error bars 
in the region where they overlap, again giving credit in the results.

We collect data from the literature.
The SL analysis by Limousin et al. (2007) yields data for the line-of-sight mass density $\Sigma(r)$  
at radii  between 3 and 271 kpc from the cluster centre \cite{limousin2007combining}.  
The 12 data $(r_i,  \Sigma_i)$ and the correlation matrix were kindly supplied by M. Limousin;
 see \cite{nieuwenhuizen2013observations} for its regularisation.
Essentially the same regime is covered by the SL analysis of Coe et al. (2010) \cite{coe2010high}, which has 20 data points;
these two data sets overlap in the window between 40 and 150 kpc with both having small errors, giving faith in the approaches.
However, discrepancies occur at small $r$: Coe et al. have 3 data points below 40 kpc while 
Limousin et al. have there 6 data points with smaller errors. Beyond 150 kpc the Coe data become increasingly noisy.
These differences stem from the different analysis and/or the neglect of correlations.
Umetsu et al. (2015) present 14 data points for $\Sigma$  between 125 kpc and 3 Mpc, derived mainly from WL  \cite{umetsu2015three},
and kindly supplied by K. Umetsu, with their correlation matrix.
These data have mostly substantially bigger errors than both previous sets, but are complementary and coincide within the error bars 
in the region where they overlap. 

Data for the WL shear $g_t$  for radii between 200 kpc and 3 Mpc have also been presented in Umetsu et al. (2015).
K. Umetsu has kindly generated for us a logarithmic binning in 13 data points in the way outlined in  \cite{umetsu2008combining}.
The resulting data agree well with the data of \cite{umetsu2008combining}, but have less scatter.

\subsection{X-ray gas}

The gas data, kindly supplied to us by A. Morandi, stem from 2 {\it CHANDRA} X-ray observations with a total exposure 
time of 150 ks, presented  in \cite{morandi2010unveiling}.
In the energy range $0.5$--$5.0$ keV the vignetting-corrected brightness image is extracted from the 2 files.
In a non-para\-metric way the gas density profile is recovered by rebinning the surface brightness into circular annuli 
and by spherical deprojection \cite{limousin2007combining}.
Fig. 1 exhibits the resulting 56 data points with $10$ kpc $<r<$ 1 Mpc.
 The S\'ersic profile $\rho=\rho_0\exp[-(r/R)^{1/n_g}]$ gave inspiration for a cored S\'ersic profile \cite{nieuwenhuizen2016dirac}, 
 
\BEQ \label{nefit}
n_e(r)=n_e^0\exp\left[k_g-k_g\left(1+\frac{r^2}{R_g^{2}}\right)^{\frac{1}{2n_g}}\right].
\EEQ
It appears to work well and has the best fit parameters  $n_e^0=0.0670 \pm  0.0028\,\cm^{-3}$, $k_g=1.98\pm 0.25$,
$R_g=21.6\pm2.7\, \kpc$ and $n_g=2.97 \pm 0.14$.  The $n_e$ error bars are asymmetric. 
The average errors of the $n_e$ lead to $\chi^2/\nu=2.08$; for the conservative maximum of upper and lower errors, $\chi^2/\nu=1.72$.
With $\nu=52$ this yields the marginally acceptable $q$-value $9\,10^{-4}$ of our spherical approximation.

Let us mention that  $n_e(r)=n_e^0(1+r^2/R^2)^{-3\beta/2}$, the popular ``$\beta$-model'', 
yields a lousy best fit, $\chi^2/\nu=6.3$.

Furthermore, data for $n_{\rm H}$ have been obtained with {\it ROSAT}/PSPC \cite{eckert2012gas}.
Given that $n_{\rm He}=n_{\rm H}/12$ when $^4$He occurs with 25\% in mass,
the connection is $n_e=7n_{\rm H}/6$.
As  $r_i$-values we take the mid of the bins and as $\Delta r_i$ their (mostly quite large) half widths.
As seen in Fig. 1, the 14 ROSAT/PSPC data have big spread in $r$, but lie at large $r$ near the cored S\'ersic profile. 
The $\chi^2$ fit is not ruined when including min$\{[n_i-n_{e}(r_i)]^2/\Delta n_{e,i}^2,[r_i-r(n_{e,i})]^2/\Delta r_i^2\}$ for its last 6 points,
where $r(n_e)$ is the inverse of $n_e(r)$ in (\ref{nefit}). Now $\nu=58$,  $\chi^2/\nu=1.69$, $q=8\,10^{-4}$.
The resulting best fit

\BEQ \label{gasfit}
&n_e^0=0.0673 \pm  0.0027\,\cm^{-3},\quad &k_g=1.90\pm 0.20,  \quad \nn\\
&R_g= 21.2\pm2.4\,\, \kpc,\qquad \quad \quad  &n_g=2.91 \pm 0.11. 
\EEQ
got  smaller errors.
For a typical $Z=0.3$ solar metal\-licity the gas mass density is $\rho_g(r)=1.167\, m_Nn_e(r)$ \cite{nieuwenhuizen2009non}.

\begin{figure}
\label{nedata}
\includegraphics[scale=0.9]{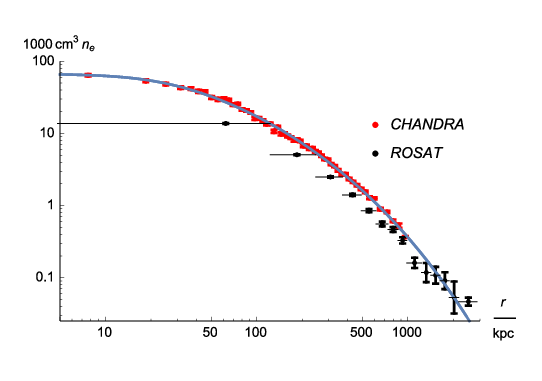}
\caption{
The  electron density $n_e$ in units of $0.001\cm^{-3}$ from \textit{CHANDRA} \cite{morandi2010unveiling} 
and \textit{ROSAT}/PSPC \cite{eckert2012gas}.
The data sets  coincide 
fairly beyond 600 kpc. Full line: the cored S\'ersic profile (\ref{nefit}).}
\end{figure}






\section{The observables}

Because the background galaxies are far removed from the cluster, the lensing effects can be thought of as occurring due to mass
projected onto the plane through the cluster centre. Hence the SL analysis yields data for the mass density accumulated along the line-of-sight,

\BEQ \label{Sigma-rho-6}
\Sigma(r)=\int_{-\infty}^\infty{\rm d} z\rho\left(\sqrt{r^2+z{}^2}\,\right).
\EEQ
In terms of the gravitational potential $\varphi$ it reads \cite{nieuwenhuizen2009non}

\BEQ \label{Sigmaphip}
\Sigma(r) =
\frac{1}{2 \pi G} \int_{0}^\infty{\rm d} s\frac{ \cosh 2s}{\sinh^2s}
\left[\varphi'(r\cosh s)-\frac{\varphi'(r)}{\cosh^2s}\right].
\EEQ

Next, one defines the $2d$ mass, that is, the mass contained in a cylinder of radius $r$ around the sight line,
$M_{2d}(r)=2\pi\int_0^r{\rm d} s\,s\Sigma(s)$ and its average over the disk, 
$\Sigmab(r)=M_{2d}(r)/\pi r^2$. This quantity can be expressed in the mass density as 

\BEQ \label{Sb1int2}
\Sigmab(r)=\frac{4}{r^2}\int_0^r\d s\,s^2\rho(s)+\int_r^\infty\d s\,\frac{4s\rho(s)}{s+\sqrt{s^2-r^2}},
\EEQ
and in terms of $\varphi$ as \cite{nieuwenhuizen2009non}

\BEQ\label{Sigmabphip}
 \Sigmab(r)=
\frac{1}{\pi G} \int_0^\infty{\rm d} s\, \varphi'(r\cosh s).
\EEQ
As imposed by their connections to $M_{2d}$, the relation $\Sigma(r)=\Sigmab(r)+\half r\Sigmab'(r)$ 
can be verified from (\ref{Sigmaphip}) and (\ref{Sigmabphip}).

In a WL analysis one determines the shear $g_t$, which relates to these quantities as

\BEQ\label{gt=}
g_t(r)=\frac{\Sigmab(r)-\Sigma(r)}{\Sigma_c-\Sigma(r)},
\EEQ
where $\Sigma_c$ is called critical density. It is fixed in the data analysis,
because the observed galaxies must be connected to a fiducial  common redshift $\bar z$.
Its values are $\Sigma_c=(0.973,0.579,0.679)\,\,\gr \,\cm^{-2}$ for the data of \cite{limousin2007combining}, \cite{coe2010high} 
and \cite{umetsu2015three}, respectively.

\subsection{Galaxies}

By definition, a galaxy cluster contains many galaxies,  which have large mutual  separations.
In the centre, the galaxy mass density is dominated by the often huge brightest cluster  galaxy (BCG, ``central galaxy'' $\cg$).
Outwards, the thermal dark matter will be dominant up to a few hundred kpc, from where on the gas density becomes relevant too.

Limousin et al. propose a BCG distribution with mass $M_\cg$, core ($co$) size $R_\co$ and extent $R_\cg$
of the form \cite{limousin2005constraining}

\BEQ\label{rho-Gal=}
\rho_\Gal(r)= \frac{M_\cg ( R_\co+R_\cg)}{2\pi^2(r^2+R_\co^2)(r^2+R_\cg^2)}.
\quad 
\EEQ
If $R_\co\ll R_\cg$ it resembles a cored isothermal distribution between these scales. Its gravitational potential reads

\BEQ
\varphi_G(r)=\frac{2GM_\cg}{\pi(R_\cg-R_\co)} &&\left( \frac{R_\co}{r}\arctan \frac{r}{R_\co} 
-\frac{R_\cg}{r}\arctan \frac{r}{R_\cg}   \right. \nn  \\ && \left. 
 +\frac{1}{2}\log\frac{1+r^2/R_\co^2}{1+r^2/R_\cg^2}  \right), 
\EEQ
where here and below we take all $\varphi=0$ at $r=0$.
There holds an explicit result for  the line-of-sight mass density,

\BEQ
\Sigma_G(r)=
 \frac{M_\cg}{2\pi(R_\cg-R_\co)} 
\left(\frac{1}{\sqrt{R_\co^2+r^2}} -\frac{1}{\sqrt{R_\cg^2+r^2}} \right) ,
\EEQ
as well as for the $2d$ mass density,

\BEQ \Sigmab_G(r) = \frac{M_\cg}{\pi r^2} 
\left(1-\frac{R_\co+R_\cg}{\sqrt{R_\cg^2+r^2}+\sqrt{R_\co^2+r^2}} \right) .
 \EEQ

\section{Newton acceleration of normal matter}
\label{sec-Newton}

The $2d$-mass, the mass inside a cylinder along the sightline of radius $r$,  relates to $\Sigmab$ as
$M_{2d}(r)=\pi r^2\Sigmab(r)$. With $M(r)\le M_{2d}(r)$ the Newton acceleration can be bounded,

\BEQ
a(r)\equiv \frac{GM(r)}{r^2}\le \frac{GM_{2d}(r)}{r^2}=\pi G\Sigmab(r).
\EEQ
While at $r=0$ one has $\Sigma(0)=\Sigmab(0)$, an isothermal decay $\rho\sim 1/r^2$ 
yields $\Sigma(r)=\half \Sigmab(r)$ at large $r$. Our interest being in the latter region,
we can use data for $\Sigma$ as a bound on $a$,

\BEQ \label{a-Sigma}
a(r)\le 2\pi G\Sigma(r). 
\EEQ
Likewise, data for the transversal shear

\BEQ
g_t(r)=\frac{\Sigmab(r)-\Sigma(r)}{\Sigma_c-\Sigma(r)}
\EEQ
can be used at large $r$, where $\Sigma(r) \ll \Sigma_c$, to estimate

\BEQ \label{a-SigmaB}
a(r)\le 2\pi G\Sigma_c g_t(r).
\EEQ
Figure 1 shows that the predictions (\ref{a-Sigma}) and (\ref{a-SigmaB}) indeed lie above the $a$ values from the fit of the NFW profile
to the $\Sigma$ data sets of Refs. \cite{coe2010high,umetsu2015three}  and the $g_t$ data \cite{umetsu2015three};
the same would happen for the \cite{limousin2007combining,umetsu2015three} and \cite{umetsu2015three} data sets.
 Figure 1 also shows the fit of our neutrino model \cite{nieuwenhuizen2016dirac} to the latter data sets.
It is seen that beyond (before) $r=200$ kpc the fits lie near (well below) the estimates (\ref{a-Sigma}) and (\ref{a-SigmaB}).

\begin{figure}\label{nedata}
\includegraphics[scale=0.9]{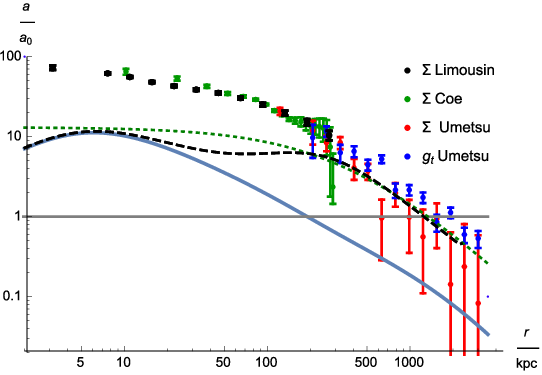}
\caption{
Full line: the Newton acceleration $a$, normalized to $a_0$, as function of the radius, as induced by the gas and galaxies in A1689.
Beyond 200 kpc it is an order of magnitude smaller than the other indicators, which points at the need for dark matter
already at $50$ kpc.
In the absence of it, modifications of Newton's law for $a<a_0$ are unlikely capable to bridge the gap.  \\
Data points: estimate (\ref{a-Sigma}) for $a$, based on data of $\Sigma$ in 
\cite{limousin2007combining,coe2010high,umetsu2015three}, and the estimate (\ref{a-SigmaB})  based on data of $g_t$ \cite{umetsu2015three}. \\
Dotted: $a$ in the optimal NFW fit (finite at $r=0$),  and
dashed: in our neutrino model, governed below 30 kpc by the BCG \cite{nieuwenhuizen2016dirac}.}
\end{figure}

\section{Newton gravity  without dark matter}

The full line in Figure 1 (the lower line) depicts the Newton acceleration due to the baryons alone, that is: the galaxies and the X-ray gas.
For $r>100$ kpc it lies an order of magnitude below the estimates  (\ref{a-Sigma}) and  (\ref{a-SigmaB}) and below the NFW and neutrino fits.
This forecasts a bad fit for Newton gravity without dark matter, the famous discovery of Zwicky.  
There seems to exist no passage to sail between the Scylla of Newton theory in the centre and the Charybdis of the need for dark matter
somewhat out of the centre.

So let us quantify this and study the  minimisation of $\chi^2$.  
While the X-ray data are confined by observations, the parameters of the BCG profile are obtained from a best fit. 
If we fix the BCG mass at a typical value of $M_{cg}=(3-4)10^{13}M_\odot$ and optimise over $R_\co$ and $R_\cg$,
 we get $\chi^2/\nu$ values of order 80-100, a failing fit.
When we let $M_\cg$ free, this has the danger of run off to a ``best'' fit with unphysical parameters.
If we look for the optimal fit we can achieve even the very good $\chi^2/\nu\sim 1.2$ when fitting the SL data of Coe with the SL and WL data of Umetsu,
but this is suspicious, since the sharper SL data of Limousin combined with the SL and WL data of Umetsu would lead to a bad value $\sim 4$.
Anyhow the fit brings $M_\cg\sim 1-2\,10^{15}M_\odot$, basically the mass of the whole cluster.
Clearly, the BCG profile tries to model the missing dark matter. 
This, however, is unphysical and unwanted, since for that case much better modelling is achieved by 
NFW or our neutrino model  \cite{nieuwenhuizen2016dirac}.
Not surprisingly, when the inner and our scale of the BCG are kept fixed at realistic values (e.g, 5 and 130 kpc, respectively), 
$M_\cg\sim 10^{13-14}M_\odot$ stays acceptable, but the fit $\chi^2/\nu\sim 7$ is lousy.

\section{MOdified Newtonian Dynamics  (MOND)}

To solve the conundrum of flat rotation curves, it has been proposed that instead of dark matter to exist,
Newton's law should be modified in the regime of small acceleration. M. Milgrom proposes to take $g=\sqrt{a_0g_m}$ for $g_m<a_0$, 
where $a_0$ is some scale of order $cH_0/2\pi$. 
When $g_m$ is written as $GM_B(r)/r^2$ with $M_B(r)$ the enclosed baryonic mass inside a sphere of radius $r$, 
$g$ will decay as $1/r$, which corresponds to a limiting rotation speed $v_\rot(\infty)=[a_0GM_B(\infty)]^{1/4}$.
Reversely, the relation $M_B(\infty)\sim v_\rot^4(\infty)$ is called the baryonic Tully-Fisher relation, and is satisfied reasonably 
well  \cite{famaey2012modified}.

It has been long realised that MOND does not function well in clusters, and that it must be augmented with neutrinos to do so
\cite{sanders2003clusters,sanders2007neutrinos}.
We shall test this in detail on the rather precise data for A1689.

In MOND the Poisson equation is modified to

\BEQ
\label{Mond1} 
\nabla\cdot\left\{\mu\left(\frac{g}{a_0}\right){\bf g}\right\}=4\pi G\rho_m
\EEQ
where ${\bf g}=\nabla\varphi$ and $g=|{\bf g}|$ and $\rho_m$ is the mass density of the normal matter,
baryons and possibly neutrinos.

\BEQ 
a_0=\frac{4cG\rho_c}{3H_0}=\frac{cH_0}{2\pi}=1.14\,10^{-10}\m \s^{-2}
\EEQ

The MOND function $\mu(y)$ goes to unity at large $y$ (large acceleration leading to Newton gravity), while $\mu=y$ at small $y$.
We adopt the most common shape 

\BEQ \mu(y)=\frac{y}{1+y}.
\EEQ

In case of spherical symmetry Eq. (\ref{Mond1}) reads

\BEQ\label{Poisson-Mond}
\mu'\left[\frac{\varphi'}{a_0 }\right] \frac{\varphi''\varphi' }{a_0 }+\mu\left[\frac{\varphi'}{a_0 }\right] 
\left(\varphi'' + \frac{2}{r} \varphi'\right)  = 4 \pi G \rho_m
\EEQ
Relating $\rho_m$ to the Newton potential $\varphi_m$ by the Poisson equation $\varphi_m''+{2}{}\varphi_m'/r=4\pi G\rho_m$,
eq. (\ref{Poisson-Mond}) can be integrated, to become $\varphi'\mu({\varphi'}/{a_0 }) =\varphi_m'$.
Inversion yields

\BEQ \label{phipMond}
\varphi'
=\frac{1}{2}\varphi_m'+\frac{1}{2}\sqrt{(\varphi_m')^2+4a_0\varphi_m' }.
\EEQ
Writing $\varphi'_m=g_m$, there result two first order equations,

\BEQ\label{gmpMond=}
\varphi'=\frac{1}{2}g_m+\frac{1}{2}\sqrt{g_m^2+4a_0g_m},
\quad g_m'+\frac{2}{r}g_m=4\pi G\rho_m. 
\EEQ

Since MOND primarily deals with the forces on massive bodies, one still has to theorise how light will propagate, 
see\cite{bekenstein2004alternative,milgrom2009bimetric}. We do not need to go that far.
Since most of the A1689 cluster will turn out to be in the Newtonian regime, and one assumes 
that MOND satisfies the same post-Newtonian corrections as General Relativity, 
light propagation is basically, or even fully, set by the equivalence principle.
With light rays moving in the gravitational potential $\varphi$, we can use
the relation (\ref{Sigmaphip}) for $\Sigma$ and (\ref{Sigmabphip}) for $\Sigmab$.

\subsection{Baryons only}

When matter consists only of baryons, Galaxies and X-ray gas, $\rho_m=\rho_G+\rho_g$ is explicit and
the equations (\ref{gmpMond=}) can be solved straightforwardly. The role of the MOND parameter
$a_0$ appears to be modest, and the fit to the data is very similar to the case of the pure Newton force,
i.e., the limit $a_0\to0$. Again, the fit is very bad, except when we allow the central galaxy profile to mimic dark matter,
in which case to fit is less bad but not acceptable.

\subsection{Adding neutrinos}

It is well  known that the MOND theory for clusters can be repaired by adding neutrinos \cite{sanders2003clusters,sanders2007neutrinos}. 
We will now consider in how far this works for the precise data of A1689.

Neutrinos, like other particles, will move in the MOND force field. 
Let us assume, for simplicity, that they have equal masses and occupations, 
with their distribution having (basically or exactly) the standard equilibrium form,

\BEQ\label{rho-nu=}
\rho_\nu(r)=\int\frac{\d^3p}{(2\pi\hbar)^3}\frac{gm_\nu}{\exp\{[\half v^2+\varphi(r)-\mu]/\sigma^2\}+1},
\EEQ
where $g$ is the number of fermionic species ($g=6$ for the active (anti)neutrinos), $m_\nu$ their common mass, $\vp=m_\nu\vv$ the 
momentum, $\vv$ the velocity, $\mu$ the chemical potential per unit mass and $\sigma$ the velocity dispersion.
In (\ref{gmpMond=}) there now occurs $\rho_m=\rho_G+\rho_g +\rho_\nu$, with the latter having the form
$\rho_\nu(r)=\rho_\nu[\varphi(r)]$. This couples the differential equations nontrivially, without changing their structure.

It appears that in the cluster the effect of $a_0$  remains very small in the presence of neutrinos.
The fit of Eqs. (\ref{gmpMond=})  with $a_0=cH_0/2\pi$  appears not to change within the error bars of the $a_0=0$ case
 of Ref. \cite{nieuwenhuizen2016dirac}. 
The physical reason is, again, that the $r>1$ Mpc region, where non-Newtonian effects become relevant,
has little statistical weight.

From the case $a_0=0$ we recall the most important results from \cite{nieuwenhuizen2016dirac}. The mass of the BCG is
\BEQ \label{m-G=}
M_G=  3.2\pm 1.0 \,10^{13}M_\odot.
\EEQ
The cluster yields a result for $gm_\nu^4$ since $\d^3p=m_\nu^3\d^3v$ in Eq. (\ref{rho-nu=}). 
The resulting mass of the particle is

\BEQ  \label{m-nu-g=}
m_\nu =1.92_{-0.16}^{+0.13} \,\left(\frac{12}{g}\right)^{1/4}
  \frac{{\rm eV}}{c^2},
\EEQ
which must refer to active and/or sterile neutrinos. 

For overdensities 500 and 200 the solution implies

\BEQ
r_{500}&=(1.7\pm0.2)\,\Mpc, \qquad M_{500}&=(1.4\pm 0.1) \, 10^{15}M_\odot ,  \nn\\
r_{200}&=(2.5\pm0.2)\, \Mpc,\qquad M_{200}&=(1.8\pm0.1)\,10^{15}M_\odot .
\EEQ

\section{MOdified Gravity (MOG)}

Also the MOdified Gravity (MOG) theory of J. Moffat attempts to do without DM \cite{moffat2005gravitational}.
The description of 106 clusters has been discussed in \cite{brownstein2006galaxy}; we single out the application to A1689.
As noted in section 2, the employed modelling of the gas density by a $\beta$-profile is bad.
At $r=0$ the value $\rho_0/1.17m_N=0.0168$/cm$^3$ is a factor 4 below our $n_e^0$ in (\ref{gasfit}) and 
the $\beta$-profile of \cite{brownstein2006galaxy} is globally rejected by the Chandra data at dozens of standard deviations.

In the cluster centre the deviation of MOG from Newton theory is non-negligible, but adjustment of the BCG parameters 
allows a reasonable fit. Far out, MOG enhances the Newton force by a factor  $1+\alpha=9.89\pm0.34$ \cite{moffat2013mog}, which works 
reasonably. But in the intermediate range between 50 and 300 kpc the data for $\Sigma$ are poorly matched.
To save the theory, an extra ingredient is needed. 
On the basis of previous section, we expect that the introduction of additional matter, 
such as neutrinos, can do the job.

\vspace{-0.5cm}

\section{Emergent Gravity (EG)}  

\vspace{-0.3cm}

\subsection{Generalities}
\vspace{-0.2cm}

E. Verlinde proposes that gravity emerges from an underlying ``polymer'' structure of space-time  \cite{verlinde2011origin}.
As discussed in a publicly available master thesis \cite{steggerda2014polymer},
one equates in this ``emergent gravity'' the gravitational field energy of dark matter inside a sphere of radius $R$
\BEQ
\frac{1}{8\pi G}\int_{r\le R}\d^3r\,(\nabla \varphi_d)^2
\EEQ
to the enclosed mass of normal matter, $M_m(R)$, times $c$ times the expansion speed at its border, $H_0R/4\pi$.
Inserting $\nabla \varphi_d=GM_d(r){\bf r}/r^3$ one gets
\BEQ
\int_0^R\d r\,\frac{GM_d^2(r)}{2r^2}=M_m(R)\frac{cH_0R}{4\pi}.
\EEQ
Taking the  derivative with respect to $R$ yields
\BEQ \label{vLM1}
\frac{GM_d^2(r)}{r^2}=M_m(r)\frac{cH_0}{2\pi}\left(1+\frac{\d \log M_m(r)}{\d\log r}\right),
\EEQ
which is equivalent to
\BEQ
\label{vLDM} 
M_d(r)=r\sqrt{\frac{a_0}{G}[M_m(r)+4\pi r^3\rho_m(r)]}.
\EEQ

\vspace{-0.5cm}

\subsection{Lensing test}
\vspace{-0.5cm}

We first express the above in a more suitable form. With $g_m=\varphi_m'=GM_m/r^2$ and $\varphi_d'=GM_d/r^2$,  Eq. (\ref{vLDM}) 
implies that the total EG potential $\varphi=\varphi_m+\varphi_d$  and $g_m$ satisfy
\BEQ
\varphi'=g_m+\sqrt{a_0(g_m+4\pi Gr\rho_m )}, \,\,\, g_m'+\frac{2}{r}g_m=4\pi G\rho_m,
\EEQ
to be compared to the expressions (\ref{gmpMond=}) for MOND.

Due to the equivalence principle, light rays will move in the gravitational potential $\varphi$, 
so that Eqs. (\ref{Sigmaphip}) and (\ref{Sigmabphip}) are valid. Neutrinos, if present, do the same, therefore
Eq. (\ref{rho-nu=}) is well motivated.

The fit to the data is as in the MOND case, because also in the EG case, the role of $a_0$ is negligible in the A1689 cluster.
In the case where matter consists only of baryons, the fit does not seriously differ from the one with the pure Newton force,
and suffers from the severe problems explained in section \ref{sec-Newton} and summarised in figure 2.

If neutrinos are present, a very good fit can be achieved, not differing within the error bars
from the neutrino fit in Newton theory or MOND theory. The results are essentially the same as in eqs. 
(\ref{m-G=}) and (\ref{m-nu-g=}).

\vspace{-0.5cm}

\section{Gravitational potential in $\small{f}(R)$ theories}
\vspace{-0.2cm}

In a modification of general relativity one  replaces in the Hilbert-Einstein action the curvature scalar $R$
by some function $f(R)$, hence the name of these theories. On the basis of this, a modification of Newton's law has
been proposed, where the gravitational potential $\varphi$ deviates from the Newton potential $\varphi_m$ in 
the form \cite{demartino2014constraining}
\BEQ \label{PoissonfR}
\varphi(r)=\frac{1+\delta e^{-r/R}}{1+\delta}\varphi_m(r),\quad
\varphi_m''+\frac{2}{r}\varphi_m'=4\pi G\rho_m,
\EEQ
where $\rho_m$ is the matter density.
We consider the interval  $-1<\delta<0$ where this enhances $\varphi_m$ at $r\gg R$, allowing to mimic dark matter effects.
Indeed, a value  $\delta\simeq-0.98$ is employed to fit cosmological data \cite{demartino2014constraining}.
But the derivative
\BEQ \label{PoissonfRp}
\varphi'(r)=\frac{1+\delta e^{-r/R}}{1+\delta}\varphi_m'(r)-\frac{\delta e^{-r/R}}{(1+\delta)R}\varphi_m(r),
\EEQ
involves both $\varphi_m'>0$ and $\varphi_m<0$, so for $-1<\delta<0$ the second term
counteracts the enhanced gravity one searches for, and even exhibits an ``anti-gravity'' effect:
at small $r,$ where $\varphi'_m(r)=4\pi G\rho_m(0)r/3$ and $\varphi_m(0)<0$, 
Eq. (\ref{PoissonfRp}) involves a domain with $\varphi'<0$. One may even wonder whether the second term in   
(\ref{PoissonfRp})  is admissible on principle grounds. Indeed,  rewriting (\ref{Sigmabphip}) as
\BEQ\label{Sigmabphip2}
 \Sigmab(r)=
\frac{1}{\pi G} \int_r^\infty\frac{{\rm d} u}{\sqrt{u^2-r^2}}\, \varphi'(u),
\EEQ
it is seen that $\Sigmab(r)\sim [ \varphi_m(0)\delta/\pi GR(1+\delta)]\log r$  goes to minus infinity for 
$r\to0$ when $-1<\delta<0$. Because of the relation $\Sigma=\Sigmab+\half r \Sigmab'$, 
which holds between (\ref{Sigmaphip}) and (\ref{Sigmabphip}) for any smooth $\varphi(r)$, it follows that also 
$\Sigma(r)$ becomes negative at small $r$, which is excluded by the data  at many standard deviations:
numerical fits are truly bad.

However, in view of the arguments of previous sections, it is unlikely that $f(R)$ theories with $\varphi'(0)=0$ 
can achieve an acceptable match with the data of the A1689 cluster.
But it is possible that such $f(R)$ theories, which next to baryons, also have heavy neutrinos, fare better.

\vspace{-0.3cm}

\section{Discussion}
\vspace{-0.2cm}

Modifications of Newton's law, such as MOND, MOG, Emergent Gravity and $f(R)$ theories have been (partly) motivated 
by the wish to avoid dark matter (DM). For galaxies, this works quite well, in particular for MOND.
For the A1689 cluster is appears not to work \cite{sanders2003clusters,brownstein2006galaxy,sanders2007neutrinos,natarajan2008mond}.
So while these alternative theories probably can do without Oort  (galactic) DM, they need still normal Zwicky (cluster) DM.
We have confirmed for MOND, shown for Emergent Gravity and coined for MOG and $f(R)$, that neutrinos can do the job.

It is broadly assumed that active neutrinos are very light $m_\nu\sim0.1$ eV. 
In this scenario our massive neutrinos are sterile neutrinos. 
Their free streaming may be ineffective if structure formation is nonlinear.
There is no reason to assume, as we did in \cite{nieuwenhuizen2016dirac} and here, 
that their masses are equal to each other, but our results will still be indicative:  
some handful  of species with masses in the few-eV range.

It is also possible that the active neutrinos (the standard ones) have such an eV mass. Then our results are more indicative.
Firstly, because of the small neutrino oscillation effects, all active neutrino masses are basically equal. 
Then the non-observation of neutrinoless double $\beta$-decay implies that the active $\nu$'s are
(nearly) of Dirac type \cite{nieuwenhuizen2016dirac}, so that each active neutrino has a sterile, right handed partner 
of (nearly) the same mass. Just using the Planck cosmic dark matter fraction, even though based on WIMPs
while the true DM would be neutrinos, this implies a neutrino mass $m_\nu=(1.861\pm0.016)\,{\eV}$ \cite{nieuwenhuizen2016dirac},
which will be tested soon at KATRIN \cite{ottenweinheimer2008}.

Either way, eV neutrino masses may explain the offset between dark and normal matter in A3827
and the central dark clump in A520 \cite{nieuwenhuizen2016dirac}.

Our analysis shows that, if sterile and/or active neutrinos have eV-scale masses, the A1689 data
will hardly discriminate between (non-)Newtonian gravity theories.

{\bf Acknowledgements}: we thank M. Limousin, A. Morandi, K. Umetsu and D. Eckert for supplying data.

\providecommand{\WileyBibTextsc}{}
\let\textsc\WileyBibTextsc
\providecommand{\othercit}{}
\providecommand{\jr}[1]{#1}
\providecommand{\etal}{~et~al.}


\begin{thebibliography}{[10]}

\bibitem{kapteyn1922first}
 \textsc{J.~Kapteyn} \jr{The Astrophysical Journal} \textbf{55}, 302 (1922).


\bibitem{oort1932force}
 \textsc{J.\,H. Oort} \jr{Bulletin of the Astronomical Institutes of the
  Netherlands} \textbf{6}, 249 (1932).


\bibitem{zwicky1933rotverschiebung}
 \textsc{F.~Zwicky} \jr{Helvetica Physica Acta} \textbf{6}, 110--127 (1933).


\bibitem{zwicky1937masses}
 \textsc{F.~Zwicky} \jr{The Astrophysical Journal} \textbf{86}, 217 (1937).


\bibitem{rubin1980rotational}
 \textsc{V.\,C. Rubin},  \textsc{W.\,K. Ford~Jr},  and  \textsc{N.~Thonnard}
  \jr{The Astrophysical Journal} \textbf{238}, 471--487 (1980).


\bibitem{clowe2006direct}
 \textsc{D.~Clowe},  \textsc{M.~Brada{\v{c}}},  \textsc{A.\,H. Gonzalez},
  \textsc{M.~Markevitch},  \textsc{S.\,W. Randall},  \textsc{C.~Jones},  and
  \textsc{D.~Zaritsky} \jr{The Astrophysical Journal Letters} \textbf{648}(2),
  L109 (2006).


\bibitem{hinshaw2013nine}
 \textsc{G.~Hinshaw},  \textsc{D.~Larson},  \textsc{Komatsu} \etal{} \jr{The
  Astrophysical Journal Suppl. Series} \textbf{208}(2), 19 (2013).


\bibitem{planck2015planck}
 \textsc{Planck-Collaboration} \jr{arXiv preprint arXiv:1502.01589} (2015).


\bibitem{bertone2010moment}
 \textsc{G.~Bertone} \jr{Nature} \textbf{468}(7322), 389--393 (2010).


\bibitem{hutsemekers2014alignment}
 \textsc{D.~Hutsem{\'e}kers},  \textsc{L.~Braibant},  \textsc{V.~Pelgrims},
  and  \textsc{D.~Sluse} \jr{Astronomy \& Astrophysics} \textbf{572}, A18
  (2014).


\bibitem{taylor2016alignments}
 \textsc{A.~Taylor} and  \textsc{P.~Jagannathan} \jr{Monthly Notices of the
  Royal Astronomical Society: Letters} \textbf{459}(1), L36--L40 (2016).


\bibitem{balazs2015giant}
 \textsc{L.~Bal{\'a}zs},  \textsc{Z.~Bagoly},  \textsc{J.~Hakkila},
  \textsc{I.~Horv{\'a}th},  \textsc{J.~K{\'o}bori},  \textsc{I.~R{\'a}cz},  and
   \textsc{L.~T{\'o}th} \jr{Monthly Notices of the Royal Astronomical Society}
  \textbf{452}(3), 2236--2246 (2015).


\bibitem{jee2012study}
 \textsc{M.~Jee},  \textsc{A.~Mahdavi},  \textsc{Hoekstra} \etal{} \jr{The
  Astrophysical Journal} \textbf{747}(2), 96 (2012).


\bibitem{clowe2012dark}
 \textsc{D.~Clowe},  \textsc{M.~Markevitch},  \textsc{Brada{\v{c}}} \etal{}
  \jr{The Astrophysical Journal} \textbf{758}(2), 128 (2012).


\bibitem{jee2014hubble}
 \textsc{M.\,J. Jee},  \textsc{H.~Hoekstra},  \textsc{A.~Mahdavi},  and
  \textsc{A.~Babul} \jr{The Astrophysical Journal} \textbf{783}(2), 78 (2014).


\bibitem{wang2016merging}
 \textsc{Q.~Wang},  \textsc{M.~Markevitch},  and  \textsc{S.~Giacintucci}
  \jr{arXiv preprint arXiv:1603.05232} (2016).


\bibitem{massey2015behaviour}
 \textsc{R.~Massey},  \textsc{L.~Williams},  \textsc{R.~Smit} \etal{}
  \jr{Monthly Notices of the Royal Astronomical Society} \textbf{449}(4),
  3393--3406 (2015).


\bibitem{schaller2015offsets}
 \textsc{M.~Schaller},  \textsc{A.~Robertson},  \textsc{R.~Massey},
  \textsc{R.\,G. Bower},  and  \textsc{V.\,R. Eke} \jr{Monthly Notices of the
  Royal Astronomical Society: Letters} \textbf{453}(1), L65--L69 (2015).


\bibitem{schild1996microlensing}
 \textsc{R.\,E. Schild} \jr{The Astrophysical Journal} \textbf{464}, 125
  (1996).


\bibitem{kroupa2012failures}
 \textsc{P.~Kroupa},  \textsc{M.~Pawlowski},  and  \textsc{M.~Milgrom}
  \jr{International Journal of Modern Physics D} \textbf{21}(14), 1230003
  (2012).


\bibitem{steinhardt2015impossibly}
 \textsc{C.\,L. Steinhardt},  \textsc{P.~Capak},  \textsc{D.~Masters},  and
  \textsc{J.\,S. Speagle} \jr{arXiv preprint arXiv:1506.01377} (2015).


\bibitem{bershadskii2015deterministic}
 \textsc{A.~Bershadskii} \jr{arXiv preprint arXiv:1510.01909} (2015).


\bibitem{nieuwenhuizen2009gravitational}
 \textsc{T.\,M. Nieuwenhuizen},  \textsc{C.\,H. Gibson},  and  \textsc{R.\,E.
  Schild} \jr{EPL (Europhysics Letters)} \textbf{88}(4), 49001 (2009).


\bibitem{nieuwenhuizen2010micro}
 \textsc{T.\,M. Nieuwenhuizen},  \textsc{R.\,E. Schild},  and  \textsc{C.\,H.
  Gibson} \jr{arXiv preprint arXiv:1011.2530} (2010).


\bibitem{milgrom1983modification}
 \textsc{M.~Milgrom} \jr{The Astrophysical Journal} \textbf{270}, 365--370
  (1983).


\bibitem{famaey2012modified}
 \textsc{B.~Famaey} and  \textsc{S.\,S. McGaugh} \jr{Living Reviews in
  Relativity} \textbf{15}(10), 1--159 (2012).


\bibitem{moffat2005gravitational}
 \textsc{J.~Moffat} \jr{Journal of Cosmology and Astroparticle Physics}
 \textbf{2006}(03), 004 (2006).


\bibitem{sotiriou2010f(R)}
 \textsc{T.\,P. Sotiriou} and  \textsc{V.~Faraoni} \jr{Reviews of Modern
  Physics} \textbf{82}(1), 451 (2010).


\bibitem{defelice2010f(R)}
 \textsc{A.~De~Felice} and  \textsc{S.~Tsujikawa} \jr{Living Rev. Rel}
  \textbf{13}(3), 1002--4928 (2010).

\newpage

\bibitem{verlinde2011origin}
 \textsc{E.~Verlinde} \jr{Journal of High Energy Physics} \textbf{2011}(4),
  1--27 (2011).


\bibitem{sanders2003clusters}
 \textsc{R.~Sanders} \jr{Monthly Notices of the Royal Astronomical Society}
  \textbf{342}(3), 901--908 (2003).


\bibitem{brownstein2006galaxy}
 \textsc{J.~Brownstein} and  \textsc{J.~Moffat} \jr{Monthly Notices of the
  Royal Astronomical Society} \textbf{367}(2), 527--540 (2006).


\bibitem{sanders2007neutrinos}
 \textsc{R.~Sanders} \jr{Monthly Notices of the Royal Astronomical Society}
  \textbf{380}(1), 331--338 (2007).




\bibitem{nieuwenhuizen2009non}
 \textsc{T.\,M. Nieuwenhuizen} \jr{EPL (Europhysics Letters)} \textbf{86}(5),
  59001 (2009).


\bibitem{nieuwenhuizen2011prediction}
 \textsc{T.\,M. Nieuwenhuizen} and  \textsc{A.~Morandi} \jr{arXiv preprint
  arXiv:1103.6270} (2011).


\bibitem{nieuwenhuizen2013observations}
 \textsc{T.\,M. Nieuwenhuizen} and  \textsc{A.~Morandi} \jr{Monthly Notices of
  the Royal Astronomical Society} p.\,stt1216 (2013).


\bibitem{nieuwenhuizen2016dirac}
 \textsc{T.\,M. Nieuwenhuizen} \jr{Journal of Physics: Conference Series}
  \textbf{701}, 012022 (2016).


\bibitem{limousin2007combining}
 \textsc{M.~Limousin},  \textsc{J.~Richard},  \textsc{E.~Jullo} \etal{} \jr{The
  Astrophysical Journal} \textbf{668}(2), 643 (2007).


\bibitem{coe2010high}
 \textsc{D.~Coe},  \textsc{N.~Ben{\'\i}tez},  \textsc{T.~Broadhurst},  and
  \textsc{L.\,A. Moustakas} \jr{The Astrophysical Journal} \textbf{723}(2),
  1678 (2010).


\bibitem{umetsu2015three}
 \textsc{K.~Umetsu},  \textsc{M.~Sereno},  \textsc{E.~Medezinski} \etal{}
  \jr{The Astrophysical Journal} \textbf{806}(2), 207 (2015).


\bibitem{umetsu2008combining}
 \textsc{K.~Umetsu} and  \textsc{T.~Broadhurst} \jr{The Astrophysical Journal}
  \textbf{684}(1), 177 (2008).


\bibitem{morandi2010unveiling}
 \textsc{A.~Morandi},  \textsc{K.~Pedersen},  and  \textsc{M.~Limousin} \jr{The
  Astrophysical Journal} \textbf{713}(1), 491 (2010).


\bibitem{eckert2012gas}
 \textsc{D.~Eckert},  \textsc{F.~Vazza},  \textsc{S.~Ettori} \etal{}
  \jr{Astronomy \& Astrophysics} \textbf{541}, A57 (2012).


\bibitem{limousin2005constraining}
 \textsc{M.~Limousin},  \textsc{J.\,P. Kneib},  and  \textsc{P.~Natarajan}
  \jr{Monthly Notices of the Royal Astronomical Society} \textbf{356}(1),
  309--322 (2005).


\bibitem{bekenstein2004alternative}
 \textsc{J.\,D. Bekenstein} \jr{arXiv preprint astro-ph/0412652} (2004).


\bibitem{milgrom2009bimetric}
 \textsc{M.~Milgrom} \jr{Physical Review D} \textbf{80}(12), 123536 (2009).




\bibitem{moffat2013mog}
 \textsc{J.~Moffat} and  \textsc{S.~Rahvar} \jr{Monthly Notices of the Royal
  Astronomical Society} \textbf{436}(2), 1439--1451 (2013).


\bibitem{steggerda2014polymer}
 \textsc{W.~Steggerda},
 \jr{https://esc.fnwi.uva.nl/thesis/centraal/ \\ files/f1716894420.pdf} (2014).


\bibitem{demartino2014constraining}
 \textsc{I.~De~Martino},  \textsc{M.~De~Laurentis},
  \textsc{F.~Atrio-Barandela},  and  \textsc{S.~Capozziello} \jr{Monthly
  Notices of the Royal Astronomical Society} \textbf{442}(2), 921--928 (2014).






\bibitem{natarajan2008mond}
 \textsc{P.~Natarajan} and  \textsc{H.~Zhao} \jr{Monthly Notices of the Royal
  Astronomical Society} \textbf{389}(1), 250--256 (2008).

\bibitem{ottenweinheimer2008}
 \textsc{E.\,W. Otten} and  \textsc{C.~Weinheimer} \jr{Reports on Progress in
  Physics} \textbf{71}(8), 086201 (2008).


\end{thebibliography}
\end{document}